# Potassium Isotopic Compositions of Enstatite Meteorites

Chen Zhao[1, 2], Katharina Lodders[1], Hannah Bloom[1], Heng Chen[1], Zhen Tian[1], Piers Koefoed[1], Mária K. Pető[3], and Kun Wang (王昆)[1*]

[1]Department of Earth and Planetary Sciences and McDonnell Center for the Space Sciences, Washington University in St. Louis, Campus Box 1169, One Brookings Drive, St. Louis, MO 63130, USA

[2]Faculty of Earth Sciences, China University of Geosciences, Wuhan, Hubei 430074, China

[3]Konkoly Observatory, Research Center for Astronomy and Earth Sciences, Hungarian Academy of Sciences, H-1121 Budapest, Hungary

[*]Corresponding author email: wangkun@wustl.edu

## ABSTRACT

Enstatite chondrites and aubrites are meteorites that show the closest similarities to the Earth in many isotope systems that undergo mass-independent and mass-dependent isotopic fractionations. Due to the analytical challenges to obtain high-precision K isotopic compositions in the past, potential differences in K isotopic compositions between enstatite meteorites and the Earth remained uncertain. We report the first high-precision K isotopic compositions of eight enstatite chondrites and four aubrites and find that there is a significant variation of K isotopic compositions among enstatite meteorites (from −2.34‰ to −0.18‰). However, K isotopic compositions of nearly all enstatite meteorites scatter around the Bulk Silicate Earth (BSE) value. The average K isotopic composition of the eight enstatite chondrites (−0.47 ±0.57‰) is indistinguishable from the BSE value (−0.48 ±0.03‰), thus further corroborating the isotopic similarity between Earth' building blocks and enstatite meteorite precursors. We found no correlation of K isotopic compositions with the chemical groups, petrological types, shock degrees, and terrestrial weathering conditions; however, the variation of K isotopes among enstatite meteorite can be attributed to the parent-body processing. Our sample of the main-group aubrite MIL 13004 is exceptional and has an extremely light K isotopic composition ($\delta^{41}K$=−2.34 ±0.12‰). We attribute this unique K isotopic feature to the presence of abundant djerfisherite inclusions in our sample because this K-bearing sulfide mineral is predicted to be enriched in $^{39}K$ during equilibrium exchange with silicates.

## 1. INTRODUCTION

Enstatite meteorites include undifferentiated chondrites and differentiated aubrites (Keil, 1989, 2010), both containing abundant (75-98 vol.%) enstatite (Watters and Prinz, 1979). Currently enstatite meteorites are the most reduced known meteorite group (Brearley and Jones, 1998). Due to their extremely reduced conditions of formation, nominally lithophile elements (*e.g.*, K) that mostly occur in silicate minerals can form unique metal, sulfide and nitride mineral assemblages (Keil, 1989; Krot et al., 2014). Enstatite chondrites can be further divided into two sub-groups: high-Fe EH and low-Fe EL. Each sub-group has several petrographic types (EH3-7 and EL3-7) according to the degree of their thermal metamorphism, with type 3 being the most unequilibrated and type 7 the most equilibrated. The genetic relationship between the two subgroups is still not well known. Keil (1989) suggested that they originate from different parent bodies because each group contains unique clasts not found in the other groups.

Aubrites are highly reduced, FeO-poor and brecciated (except for Shallowater) enstatite orthopyroxenites (Keil, 1989). All aubrites appear to have formed on the same parent body, which has experienced extensive melting and igneous differentiation (Keil, 1989); however the genetic relationship between the enstatite chondrites and aubrites is still under debate (Keil, 1989, 2010). Some detailed discussion on the mineralogy and trace elements in aubrites can be found in previous studies (*e.g.*, Watters and Prinz, 1979; Floss et al., 1990).

Enstatite meteorites have received attention due to their isotopic similarities to the Earth, especially for oxygen, which is the most abundant element in both the Earth and stony meteorites. Both enstatite chondrites and aubrites plot on the Terrestrial Fractionation line defined by terrestrial samples on the $\delta^{18}O$ vs. $\delta^{17}O$ plot (Clayton et al., 1984), suggesting a genetic relationship between the Earth and enstatite meteorites. As such, enstatite chondrites have been suggested as a proxy for the building blocks of the Earth (Javoy, 1995; Javoy et al., 2010). In addition to the O isotope systematics, enstatite chondrites and the Earth display indistinguishable mass-independent isotope effects such as excesses or depletions in the heavy isotopes of iron peak elements, $^{48}Ca$, $^{50}Ti$, $^{54}Cr$, $^{64}Ni$; in the *p*-process isotope $^{92}Mo$, but also in the *s*-process isotope $^{100}Ru$ (Dauphas et al., 2002; Trinquier et al., 2007; Trinquier et al., 2009; Qin et al., 2010; Burkhardt et al., 2011; Steele et al., 2012; Tang and Dauphas, 2012; Zhang et al., 2012; Dauphas et al., 2014; Fischer-Gödde et al., 2015; Mougel et al., 2018).

The isotopic compositions of Mg, Ca, Fe, Zn, and Rb which follow mass-dependent fractionations are also identical for enstatite chondrites and the Earth within currently available measurement precisions (Teng et al., 2010; Moynier et al., 2011; Wang et al., 2014; Huang and Jacobsen, 2017; Pringle and Moynier, 2017). One exception observed so far is the enrichment of light Si isotopes in enstatite meteorites when compared to the Earth (Georg et al., 2007). This could be due to the presence of isotopically light Si isotopes in the metal phase of enstatite chondrites, or to fractionation between Mg-silicates and nebular gas during condensation (Georg et al., 2007; Fitoussi et al., 2009; Fitoussi and Bourdon, 2012; Savage and Moynier, 2013;

Dauphas et al., 2015). The S stable isotope composition of enstatite chondrites is also different from that of the Earth; however, S is highly chalcophile and the difference can be attributed to isotope fractionation during core formation (Defouilloy et al., 2016).

There are several other differences between enstatite meteorites and the Earth, notably in their oxidation states and major element ratios (*e.g.*, Mg/Si and Al/Si) (Larimer and Anders, 1970; Baedecker and Wasson, 1975). As such, it has been suggested that enstatite meteorites and the Earth were derived from the same isotopic reservoir, but experienced different evolutionary paths (*e.g.*, Dauphas, 2017).

Potassium is a moderately volatile element with a 50% condensation temperature ~ 1006K at $10^{-4}$ bar in a gas of solar composition (Lodders, 2003), and is normally (i.e., with terrestrial rocks as guide) considered a lithophile, incompatible element. However, under the low oxygen fugacity ($\log(fO_2)$=FMQ−8 to −10; where FMQ is the Fayalite-Magnetite-Quartz buffer) conditions inferred for enstatite meteorites (Fogel et al., 1989; Casanova et al., 1993) the cosmochemical and geochemical character changes (Petaev and Khodakovsky, 1986). In a solar gas enriched in C-rich interplanetary dust particles (e.g., a gas with a C/O ratio ≥0.9, in contrast to the solar ratio of 0.5; Lodders, 2003) necessary to explain the reduced mineralogy of enstatite chondrites, potassium condensation includes sulfides such as djerfisherite [$K_6(Fe,Cu,Ni)_{25}S_{26}Cl$] (Ebel and Alexander, 2011; Ebel and Sack, 2013). As a result, the geochemical character of K in enstatite chondrites must be considered both lithophile and chalcophile, since djerfisherite, which is made up of about 10% K, can be found in enstatite meteorites.

Potassium has two stable isotopes: $^{41}K$ (6.7302 atom%) and $^{39}K$ (93.2581 atom%); the fraction of long-lived $^{40}K$ is less than 0.0117 atom% in the Earth and stony meteorites (Meija et al., 2016). These characteristics give K and its isotopes the potential to be used as a tracer for understanding the different evolutionary history of the enstatite meteorites and the Earth, especially for moderately volatile element accretion. The Earth is depleted in K and the other moderately volatile elements by almost an order of magnitude compared to enstatite chondrites. The K/U ratio of the Earth is 13,800 (Arevalo et al., 2009), while the K/U ratio of the enstatite chondrites is ~100,000 (Lodders and Fegley, 1998). So far only two enstatite chondrites and two aubrites have been analyzed for K isotopes: EH4 Abee and Indarch (Kempe and Zähringer, 1966; Burnett et al., 1966; Humayun and Clayton, 1995); aubrites Norton County and Peña Blanca Spring (Burnett et al., 1966). There was no detectable K isotopic difference between enstatite meteorites and Earth.

Over the past years, the analytical precision of K isotopes based on Multiple-Collector Inductively-Coupled-Plasma Mass-Spectrometer (MC-ICP-MS) has been significantly improved (Li et al., 2016; Wang and Jacobsen, 2016a; Hu et al., 2018; Morgan et al., 2018) and resolvable K isotopic differences among extraterrestrial and terrestrial samples have been observed. Therefore in order to investigate the evolution of moderately volatiles in enstatite meteorites and to compare the evolution of volatiles with that of the Earth, we measured high-precision K isotopic compositions of 12 enstatite meteorites samples. The K isotopes of enstatite chondrites and aubrites can also help illuminate the genetic relationships and thermal history of different sub-groups of enstatite meteorites and better understand the mechanisms of moderately volatile element fractionation during either nebular or planetary condensation/vaporization events.

## 2. SAMPLES AND ANALYTICAL METHODS

### 2.1 Sample description

We have studied twelve bulk samples from the three different enstatite meteorite groups: EH-, EL-chondrites and aubrites. All samples are *finds* from Antarctica. The acronyms in Antarctic meteorites' names represent the places where they were found (e.g., ALH, EET, GRO, LAP, LAR, LON, MAC, MET, MIL, PCA, PRA, PRE, QUE, RBT, RKP represent Allan Hills, Elephant Moraine, Grosvenor Mountains, LaPaz Icefield, Larkman Nunatak, Lonewolf Nunatak, MacAlpine Hills, Meteorite Hills, Miller Range, Pecora Escarpment, Mount Pratt, Mount Prestrud, Queen Alexandra, Roberts Massif, Reckling Peak respectively). A previous detailed study compared Antarctic *finds* with meteorite *falls* and showed that the K isotopic compositions of Antarctic meteorites have not been altered compared to meteorite falls (Tian et al., 2018). The eight enstatite chondrites analyzed include three EH3 (LAR 12252, LAR 06252 and MIL 07028), one EH4 (EET 96135), one EL3 (MAC 88136), one EL4 (MAC 02747) and two EL6 (ALHA81021 and LON 94100). MAC 88136 is a regolith breccia, while MAC 02747, ALHA81021, LON 94100, LAR 06252, and EET 96135 are unbrecciated (Rubin, 2015). The brecciation characteristics of MIL 07028 and LAR 12252 are not reported. Four main-group aubrite samples were analyzed (ALHA78113, ALH 84009, MIL 13004 and LAR 04316). All aubrites are brecciated: LAR 04316 is a regolith breccia, ALHA78113 is a polymict breccia and the remaining aubrites are monomict breccias (Rubin et al., 1997; Keil, 2010; Rubin, 2010, 2015). To test the reproducibility of the measurements, 6 enstatite chondrite analyses were replicated (labeled as #1 and #2 in **Table 1**) from the same homogenized powders. Possible K isotopic heterogeneity within bulk aubrite samples was tested by analyzing two fragments from each of the three aubrites (ALHA78113, MIL 13004 and LAR 04316; labeled as fragment 1 and fragment 2 in **Table 2**).

### 2.2 Analytical methods

Meteorite samples were crushed into fine powders in an agate mortar. Aliquots (~100mg) were then dissolved under pressure in Parr digestion vessels in concentrated $HF/HNO_3$ mixture (7ml HF+ 2ml $HNO_3$) at 180°C for 48h to ensure a total dissolution and dried under heat lamps. Samples were then redissolved with 6 N HCl for another 24h.

Potassium was separated by cation-exchange chromatography using the procedures described in Chen et al. (2019). Briefly, the samples were dissolved in 0.7 N $HNO_3$, centrifuged and then loaded into chromatography columns (ID=1.5cm; filled with 17 mL Bio-Rad AG50W-X8 100-200 mesh cation-exchange resin). Matrix elements were eluted with 0.7 N $HNO_3$ and then K was collected. To further purify K from matrix elements, we repeated the same procedure again with a smaller column (ID=0.5cm; filled with 2.4 mL resin). The yield of the procedure was monitored at all times by collecting elute pre-cuts and post-cuts (collected before and after the K-cuts) with total K yields all > 99%.

Samples were analyzed with a Thermo Scientific Neptune *Plus* MC-ICP-MS at Washington University in St. Louis. Instrumental mass bias was corrected by using the sample-standard bracketing protocol. The K isotopic compositions are reported in the delta notation, where $\delta^{41}K = ([(^{41}K/^{39}K)_{sample} / (^{41}K/^{39}K)_{standard} - 1] \times 1000)$. The standard used here is NIST SRM 3141a. One geostandard (BHVO-2) was always analyzed in the same session as an external standard for quality control. The internal (within-run) reproducibility was typically ~0.05 ‰ (95% confidence interval; 2σ). The long-term (~20 months) reproducibility of this method has been evaluated at 0.11‰ (2SD; Chen et al., 2019). The total procedural blank of this method is 0.26 μg (Chen et al., 2019) and is negligible for all the samples here.

## 3. RESULTS

Potassium isotopic compositions are reported in **Table 1** and **2** for bulk enstatite chondrites and aubrites, respectively. All uncertainties are quoted as 95% confidence intervals, 2σ. The geostandard BHVO-2 ($\delta^{41}K$ = −0.46 ±0.03‰; n=91) is in excellent agreement with previously published values for BHVO (Li et al., 2016; Wang and Jacobsen, 2016b; Wang and Jacobsen, 2016a; Hu et al., 2018; Morgan et al., 2018).

### 3.1. Potassium isotopic compositions of enstatite chondrites

As shown in **Table 1** and **Fig. 1**, the K isotopic compositions of enstatite chondrites range from −0.92 ±0.04‰ (EL6 ALHA81021) to −0.18 ±0.03‰ (EL6 LON 94100) and scatter around the Bulk Silicate Earth (BSE) value (−0.48 ±0.03‰; Wang and Jacobsen, 2016a). The error-weighted average of all enstatite chondrites is −0.47 ±0.57‰ (2SD, n=8), which is indistinguishable from the BSE value.

The K isotope data for the three EH3 chondrites range from $\delta^{41}K$ = −0.61 ±0.02‰ to −0.18 ±0.03‰ with an average of −0.41 ±0.43‰ (2SD, n=3). The EH4/5 chondrite EET 96135 shows a $\delta^{41}K$ of −0.33 ±0.02‰, which is in agreement with the EH4 sample (Indarch; +0.3 ±0.8‰) reported by Humayun and Clayton (1995). The K isotopic composition of the EL3 chondrite MAC 88136 ($\delta^{41}K$ = −0.74 ±0.05‰) is essentially identical to the value of the EL4 chondrite MAC 02747 ($\delta^{41}K$ = −0.76 ±0.02‰). The EL6-chondrites show considerable variability, with $\delta^{41}K$ ranging from −0.92 ±0.04‰ to −0.18 ±0.03‰. The average $\delta^{41}K$ of EL6 chondrites is −0.43 ±1.04‰ (2SD, n=2). Six analyses were repeated using aliquots from the same powdered samples through the full procedure from sample digestion, to column purification, to MC-ICP-MS analysis. All replicate analyses agree within our typical internal reproducibility (~0.05‰; see **Table 1** and **Fig. 2**).

### 3.2. Potassium isotopic compositions of aubrites

The four aubrites display a large range in K isotopic compositions (−2.34‰ < $\delta^{41}K$ < −0.37‰; see **Table 2** and **Fig. 1**). Excluding MIL 13004, the average value of aubrites is $\delta^{41}$K = −0.55 ±0.39‰ (2SD, n=3), which is indistinguishable from the average value of −0.47 ±0.03‰ for all enstatite chondrites and that of the Bulk Silicate Earth (−0.48 ±0.03‰; Wang and Jacobsen, 2016a). Because all aubrites used in this study are breccias, we measured multiple different fragments from each meteorite in order to confirm the data and check the homogeneity of the K isotopic distribution within the sample (**Table 2** and **Fig. 2**). The two fragments of ALHA78113 (−0.63 ±0.05‰ vs. −0.63 ±0.05‰) and the two fragments of MIL 13004 (−2.34 ±0.04‰ vs. −2.36 ±0.12‰) have identical K isotopic compositions within errors. The $\delta^{41}$K of the two LAR 04316 fragments (−0.34 ±0.04‰ and −0.54 ±0.08‰) indicating some possible variability but the difference is still within our long-term reproducibility (±0.11‰; Chen et al., 2019). These results suggest homogeneity of the K isotopic composition within ~0.1‰ for aubrites at the sub-centimeter-scale.

## 4. DISCUSSION

### 4.1. The K isotopic composition of enstatite chondrites compared to the Bulk Silicate Earth.

Previous K isotope studies found no resolvable difference (within ~0.5‰) between enstatite meteorites and the Earth based on two enstatite chondrites and two aubrites (Kempe et al., 1966; Burnett et al. 1966; Humayun and Clayton, 1995).

We analyzed 8 EC samples that cover nearly all petrological and chemical types of enstatite chondrites. The improved analytical precision (~0.05‰) enabled us to resolve K isotopic differences between the Bulk Silicate Earth (−0.48 ±0.03‰; Wang and Jacobsen, 2016a) and individual enstatite chondrites for the first time (varying from −0.92 to −0.18‰). We observe resolvable K isotope fractionation at the 0.5‰ levels (*i.e.*, best analytical precision in Humayun and Clayton (1995) between individual enstatite chondrite samples. However, it is important to note that the K isotopic compositions of all enstatite chondrites scatter around that of the BSE and the average value calculated from all enstatite chondrites ($\delta^{41}$K = −0.47 ±0.57‰; 2SD, n=8) is identical to the BSE value (−0.48 ±0.03‰; Wang and Jacobsen, 2016a). Even for the EL6 chondrites, the type that shows the largest variation, the average $\delta^{41}$K value (−0.43 ±1.04‰; 2SD, n=2) is indistinguishable from that of the BSE.

As shown in **Fig. 3** and **Fig. 4**, compared to carbonaceous and ordinary chondrites (Bloom et al., 2018; Ku and Jacobsen, 2019), enstatite chondrites are the only group with an average K isotopic composition identical to that of the BSE. In contrast, most carbonaceous chondrites (averaging at ~−0.28‰) display heavier K isotopic compositions than the BSE (−0.48‰), while most ordinary chondrites (averaging at ~−0.70‰) are isotopically lighter (Bloom et al., 2018; Ku and Jacobsen, 2019). Tian et al. (2018) reported K isotopic compositions for lunar, Martian, and eucrite meteorites, which are all enriched in $^{41}$K compared to the BSE. Among all undifferentiated and differentiated meteorites that have been analyzed so far, enstatite chondrites are the only meteorites that have the same average K isotopic composition as the Earth (See **Fig. 3** and **Fig. 4**). The high-precision K isotopic composition of enstatite chondrites is consistent with other stable isotope systems (*e.g.*, Mg, Ca, Fe, Zn, and Rb) and mass-independent isotope systems (*e.g.*, $^{48}$Ca, $^{50}$Ti, $^{54}$Cr, $^{64}$Ni, $^{92}$Mo, and $^{100}$Ru anomalies) (Dauphas et al., 2002; Trinquier et al., 2007; Trinquier et al., 2009; Qin et al., 2010; Teng et al., 2010; Burkhardt et al., 2011; Moynier et al., 2011; Steele et al., 2012; Tang and Dauphas, 2012; Zhang et al., 2012; Dauphas et al., 2014; Wang et al., 2014; Fischer-Gödde et al., 2015; Huang and Jacobsen, 2017; Pringle and Moynier, 2017; Mougel et al., 2018), which overall could suggest a kinship of the matter that accreted to the Earth and enstatite meteorites.

### 4.2. The origin of the K isotopic fractionation in enstatite chondrites

The enstatite chondrites analyzed in this study display significant $\delta^{41}$K variations (−0.92 to −0.18‰) without definitive trends among the chemical groups (EH and EL-chondrites), petrological types (3-6), shock degrees (S2-S4) and terrestrial weathering conditions (see **Fig. 5**). To understand the possible mechanism(s) accounting for the K isotopic variations among enstatite chondrites, we consider the following four possibilities: 1) terrestrial contamination and weathering; 2) spallogenic and cosmogenic effects; 3) parent-body thermal processing; and 4) impact vaporization.

*4.2.1 Terrestrial contamination and weathering*

All meteorites in this study are "finds" from Antarctica and their K isotopic compositions could potentially be altered by terrestrial contamination. In general, the terrestrial ages of Antarctic meteorites vary from 2000 to 1 million years (Nishiizumi et al., 1989; Jull et al., 1998). Among all meteorites in this study, only ALH 84007 (paring with ALH 84009) has been studied for its terrestrial age. Its terrestrial age of 2.4-4.6 ± 1.4 ka is among the shortest terrestrial ages reported for Antarctic meteorites (Jull et al., 1998). The weathering processes of meteorites commonly involve the formation of rust, oxidation of metal, hydrolysis of silicates, hydration, carbonation, and dissolution (Bland et al., 2006). The presence of evaporite minerals on the surface could serve as an indicator for extensive hydrous alteration (*i.e.*, mobilization of the soluble elements like K; Velbel et al., 1991). Among the eight EC samples in this study, only two (LON 94100 and LAR 06252) contain observable evaporite minerals (see **Table 1**; letter "e" represents evaporite mineral visible). As shown in **Table 1** and **Fig. 5a**, when compared to the other six samples these two evaporite-bearing samples (LON 94100 and LAR 06252) do not show more extreme $\delta^{41}$K values. For example, EL6 chondrite LON 94100 (weathering grade Be; moderately weathered) has a heavier isotopic composition than EL6 chondrite ALHA81021 (grade A; minor weathering); in contrast EH3 chondrite LAR 06252 (grade Be; moderately weathered) shows a lighter isotopic composition than EH3 chondrite LAR 12252 (grade A; minor weathered). Hence no correlation between the degree of terrestrial weathering and the K isotopic composition could be found among Antarctic enstatite chondrites. We want to note, however, that the assigned degree of terrestrial weathering only describes the *overall* weathering conditions of the recovered meteorites. It is possible that the specific sample fragments analyzed here represent a different weathering condition compared to the bulk, which might explain why no correlation is seen between degree of terrestrial weathering and K isotopic composition.

Additional evidence that terrestrial weathering cannot be a viable explanation for the K isotopic variations observed in enstatite chondrites is given in Tian et al. (2018). They compared K isotopic compositions of eucrite falls and finds from Antarctica, as well as meteorite finds from the desert in North West Africa (NWA). Antarctic meteorites (finds) showed identical K isotopic compositions to meteorite falls, while NWA finds showed a clear deviation from the values of meteorite falls with K isotopic compositions close to those of terrestrial samples. This observation is consistent with previous studies that show Antarctic meteorites are less prone to weathering due to the extremely cold and dry climate (*e.g.* Crozaz and Wadhwa, 2001). Furthermore, compared to eucrite meteorites, enstatite chondrites generally have similar or even higher abundances of K (*e.g.*, Kitts and Lodders, 1998), so the addition (if any) of terrestrial K would have to be larger to affect the K isotopes of Antarctic enstatite meteorites.

In summary, terrestrial contamination and weathering cannot explain the K isotopic variation observed in enstatite chondrites. However, we are also aware that this must be experimentally verified by measurements of K isotopes in samples from observed enstatite meteorite falls, which is planned in the future once such material will be made available by curators.

*4.2.2 Spallogenic and cosmogenic effects*

Since K has only two stable isotopes ($^{39}$K and $^{41}$K), and its long-lived radioactive isotope ($^{40}$K) is usually far less abundant and cannot be measured with the current method due to the direct interference from $^{40}$Ar, it is difficult to distinguish mass-dependent isotopic fractionation effects on the $^{41}$K/$^{39}$K ratio from potential changes of the $^{41}$K/$^{39}$K ratio due to spallogenic and cosmogenic effects. Solar irradiation and galactic cosmic ray (GCR) radiation effects for enstatite chondrites and aubrites are well known. GCR irradiation effects are noticeable in the noble gas record of enstatite chondrites and aubrites (*e.g.*, Lee et al., 2009). Moreover, secondary neutrons from spallation reactions thrive in Fe-poor samples and thus neutron-induced isotope shifts of the Sm and Gd isotopic compositions are exemplary in the Norton County aubrite (*e.g.*, Hidaka et al., 1999). On average, aubrites have the largest cosmic ray exposure ages among stony meteorites (Lorenzetti et al., 2003; Herzog and Caffee, 2014).

Therefore consideration of irradiation effects on the K isotopic composition should be relevant here. Cosmic ray bombardment is known to produce all isotopes of K (and other lighter stable and radioactive isotopes such as $^{36}$Cl, and Ar isotopes) by spallation of heavier elements (in particular abundant Fe) in meteorites. Production of K has been extensively analyzed in iron meteorites (no indigenous K) where $\delta^{41}$K > 1,000‰ was frequently found (Voshage et al., 1983). Production of K isotopes from Fe should be most noticeable in samples with high Fe contents (more target nuclei) and low total K (less dilution of cosmogenic K by indigenous K). The EH chondrites are very Fe-rich (bulk Fe content 30.5 wt.%; Lodders and Fegley, 1998) and given similar exposure ages, spallation of Fe would lead to a larger change in their $^{41}$K/$^{39}$K ratios than in EL-chondrites (24.8 wt.% Fe on average; Lodders and Fegley, 1998) and aubrites (0.41 wt.% Fe on average; Keil, 2010). However, as shown in **Fig. 1**, EH-EL chondrite and aubrites span the same range in K isotopes. In addition, there is no hint of any trend between $\delta^{41}$K and the Fe/K for individual samples (see **Fig. 6a**).

Radioactive $^{41}$Ca (half-life: 0.1 Ma) can be produced by thermal neutron-capture through $^{40}$Ca(n, $\gamma$)$^{41}$Ca and this secondary cosmic ray irradiation product has been measured in several aubrites (Herzog et al., 2011; Welten et al., 2004). For long irradiation durations (>> half life of $^{41}$Ca) and high neutron fluxes, decay of $^{41}$Ca to $^{41}$K may lead to measureable effects in very Ca-rich samples of aubrites. This effect is pronounced in large meteorites (such as the Norton County aubrite) since the thermalization of neutrons occurs deep in the meteorite. Thus, knowing the meteorite size in space and the location of the samples analyzed within meteorites becomes important and must be considered in the future. As shown in **Fig. 6b**, however, there is no trend between $\delta^{41}$K and the Ca/K for individual samples.

Since spallation and neutron capture yields also depend on the exposure time there should be a correlation between the meteorites' K isotopic compositions and their exposure ages. Patzer and Schultz (2001) reported cosmic-ray exposure ages for EL-chondrites MAC 88136 and LON 94100, and EH-chondrite EET 96135. Lorenzetti et al. (2003) reported cosmic-ray exposure ages for aubrites ALHA78113 and ALH 84008 (paring with ALH 84009 in this study). As shown in **Fig. 6c**, although the data are limited (one or two samples from every group), there is a slight hint of correlation between $\delta^{41}$K and the cosmic-ray exposure ages within each studied group, which suggests that some of the K isotopic variation observed in enstatite chondrites may be influenced by spallogenic and cosmogenic effects. Nevertheless, a larger data set is needed to confirm the proposed correlation.

*4.2.3 Parent-body thermal processing*

Significant K isotopic variations are observed in the eight enstatite chondrites studied here and parent-body processing could be another possible explanation for such variations. Except for the rare type 7, samples in this study cover the complete spectrum of petrological types: three EH3, one EH4/5, one EL3, one EL4 and two EL6. In **Fig. 5b**, we plotted the petrological type vs. the K isotopic compositions for EH and EL chondrites. The least metamorphosed samples (type 3-4) from EL chondrites are more enriched in lighter K isotopes than observed in EH chondrites (**Fig. 5b**). Since it has been shown that different parent bodies have unique K isotope compositions due to different volatilization histories (Wang and Jacobsen, 2016b; Tian et al., 2018), the difference between unequilibrated EH and EL chondrites is consistent with the notion that they originated on different parent bodies with distinct volatile depletion histories (Keil, 1989)

There is no obvious trend in K isotopic composition with petrologic type for EH chondrites, and unequilibrated EH3/4 chondrites span a wide range in K isotopic compositions. There could be a trend towards lighter or heavier K isotopic compositions with petrologic type for EL chondrites because we have two EL6 chondrites (LON94100 and ALHA81021) with a large difference in isotopic compositions ($-0.92‰ < \delta^{41}K < -0.18‰$). As discussed above and below, the difference in K-isotopic composition of our EL6 chondrites is difficult to explain by weathering or shock effects. Until more EL chondrites are analyzed, conclusions about any possible trends of isotopic variations with petrologic types are premature.

*4.2.4 Impact vaporization*

One mechanism for the relatively large variations in K isotopic compositions observed for EH3 and EL6 chondrites is evaporation-driven kinetic K isotopic fractionation generated by impacts. The EL6 chondrites experienced thermal metamorphism and impact melting and/or brecciation (Rubin et al., 2009). Rubin and Wasson (2011) suggest that ~60% of EL6 meteorites were formed by impact melting. Kinetic isotopic fractionation during vaporization would enrich EL6 samples in the heavier isotopes if they represent evaporation residues. Enrichments in heavy isotopes have been observed for other moderately volatile elements such as the Cd and Zn (Wombacher et al., 2003; Wombacher et al., 2008; Moynier et al., 2011). However, we do not find a correlation between impact shock stage level (**Table 1**) and $\delta^{41}$K (**Fig. 5c**) and our two EL6 chondrites are formally both classified as shock stage S2. In addition, the one sample in this study classified as a breccia, EL3 MAC 88136, has the same K isotope composition (−0.74 ±0.05‰ vs. −0.76 ±0.02‰) as the unbrecciated samples (e.g., EL4 MAC 02747). As such, impact brecciation does not appear to have fractionated K isotope compositions of enstatite chondrites.

Although no correlation between impact shock level and $\delta^{41}$K is seen, the currently observed shock levels of EL6 chondrites may not represent the real degree of their impact-shock levels as they also experienced annealing after impact(s) (Rubin et al., 1997; Rubin, 2015). Therefore, impact evaporation-driven kinetic K isotopic fractionation remains one plausible explanation for the large $\delta^{41}$K variations observed for EL6 meteorites. Udry et al. (2019) recently reclassified four meteorites (NWA 4799, 7214, 7809, 11071) from aubrites to enstatite chondrite impact melts based on multiple lines of petrological and geochemical evidences. Currently, no enstatite chondrite impact melts have been analyzed in K isotopes. If the impact evaporation-driven kinetic K isotopic fractionation discussed above indeed generated the K isotope variation observed in EL6, we would expect heavy K isotope enrichments in the enstatite chondrite impact melts since they have experienced intensive impact melting and evaporation. Such a hypothesis could be tested on those impact melts in the future to investigate the impact evaporation-driven kinetic K isotopic fractionation mechanism proposed here. At last, Boyet et al. (2018) proposed that EL3 and EL6 chondrites are originated from two different parent bodies based on Nd isotopes. The K isotope difference between EL3 and EL6 observed here is consistent with this proposal and indicates that the two parent bodies experienced different impact histories.

The variations in K isotopes among EH chondrites are difficult to relate to shock stage because only two EH chondrites have a shock level assigned to them (**Table 1** and **Fig. 5c**). We suspect that the K isotopic composition of LAR 12252 (EH3, no formal shock stage known) could be affected by impact vaporization. Compared to other EH3 chondrites, this meteorite is depleted in K (220-214 ppm vs. 498-668ppm) and enriched in heavy K isotopes (−0.18‰ vs. −0.61‰ to −0.35‰), which is consistent with K isotopic fractionation due to impact vaporization.

**4.3. The origin of the K isotopic fractionation in aubrites**

Aubrites are all breccias with the exception of Shallowater. In contrast to enstatite chondrites, aubrites are depleted in metal and sulfide phases indicating metal-sulfide segregation on the aubrite parent asteroid. It was suggested that their parent body was involved in a collision sufficiently powerful to result in global disruption, and that the material was subsequently reassembled, to form a "rubble pile" asteroid (Wolf et al., 1983). As shown in **Table 1** and **Fig. 1** if MIL 13004 is excluded, the average $\delta^{41}K = -0.55 \pm 0.06‰$ (2SD, n=3) is indistinguishable from average value of enstatite chondrites as well as the BSE value. The K concentrations in aubrites are strongly depleted compared to both EH and EL chondrites; however aubrites have similar isotopic compositions to those of EH and EL chondrites. This is opposite to what would be expected if the K depletion were due to impact vaporization or the degassing of the aubrite parent-body magma ocean, assuming the aubrites started with an enstatite-chondrite-like K concentration and isotopic composition, which is a plausible assumption considering the identical O isotopic compositions of aubrites and enstatite chondrites. Importantly, our high precision data clearly shows that the process responsible for depleting K in aubrites did not cause any significant K isotopic fractionation relative to enstatite chondrites. Thus the K depletion in aubrites is more likely the result of magmatic differentiation processes rather than the magma-ocean degassing.

An outlier among the aubrites is MIL 13004, which has the lightest K isotopic composition ($\delta^{41}$K = −2.34 ± 0.04‰) among all extraterrestrial samples analyzed so far (*e.g.*, Tian et al., 2018; Bloom et al., 2018; Wang and Jacobson, 2016b; Ku and Jacobsen, 2019). We repeated the measurement of a different fragment of the same chip of this meteorite, and obtained the same result (−2.36 ±0.12‰ vs. −2.34 ± 0.04‰; see **Fig. 2**). This large depletion in $^{41}$K is hard to explain by any known processes. The largest variation of K isotopes seen among terrestrial natural samples is only ~1‰, and this is due to low-temperature processes such as hydrothermal alteration and clay formation (Parendo et al., 2017; Morgan et al., 2018; Santiago Ramos et al., 2018). Such a large K isotopic fractionations has not yet been observed among carbonaceous and ordinary chondrites or differentiated lunar, martian and eucrite meteorites (Bloom et al., 2018; Tian et al., 2018; Ku and Jacobsen, 2019). Interestingly, other moderately volatile elements such as Zn also show an extreme enrichment of lighter isotopes in aubrites compared to enstatite chondrites (Moynier et al., 2011). This enrichment of lighter Zn isotopes has been interpreted as the back-condensation by the vapor phase, which is enriched in lighter Zn isotopes. The vapor phase is caused by the liberation of volatiles in EL6 chondrites during impacts causing the EL6 to become enriched in heavy Zn isotopes (Moynier et al., 2011). This mechanism originally proposed for Zn isotopes in aubrites can possibly explain the isotopically light K in MIL 13004. The K concentration in MIL 13004 (313-340 ppm) is higher than the majority of aubrites, which supports the hypothesis that isotopically light K condensed back from a vapor phase.

Alternatively, the uniquely low $\delta^{41}$K value observed in MIL 13004 could be also attributed to inter-mineral K isotope fractionation in aubrites. This may be characteristic to enstatite meteorites due to the extreme reducing conditions of formation. Under such conditions typical lithophile elements (*e.g.*, K) exhibit chalcophile behavior and can form unique K-bearing sulfide phases. Potassium in aubrites has been found in minerals such as enstatite (<90ppm), albite (4730ppm), alabandite (<1570 ppm) and djerfisherite (~ $10^5$ppm) (data from the Peña Blanca Spring aubrite from Lodders et al., 1993). Djerfisherite is an accessory mineral and its distribution in aubrites is highly heterogeneous. Mass-balance calculations indicate that, djerfisherite could be the main contributor (>90%) of the K inventory even if the djerfisherite takes up only 1% of the total mass of the aubrite fragment (MIL 13004) studied. In djerfisherite the nearest neighbors of K atoms are Cl and S. The bond lengths of K-Cl and K-S are much larger than the K-O bond lengths in silicates.

According to isotopic fractionation theory (*e.g.*, Schauble, 2004) and *ab initio* calculation, djerfisherite should be significantly enriched in lighter K isotopes when in equilibrium with K-bearing silicate minerals (Li et al., in review). Thus the equilibrium K isotopic fractionation between mineral phases (K-bearing silicates and sulfides) in aubrites *could* account for the higher concentration of K and the isotopically light composition in MIL 13004 (see **Fig. 7**). However, there is no detailed mineralogy and petrology study yet on MIL 13004 beyond initial classification. Future work needs to be done to test such hypothesis by isolating djerfisherite from aubrites and analyzing its K isotopic composition.

.

## 5. CONCLUSION

We have studied the K isotopic compositions of 12 enstatite chondrites and aubrites in various chemical groups (EH and EL chondrites, and main-group aubrites), petrological types (3-6), shock degrees (S1-S4), terrestrial weathering conditions, and cosmic-ray exposure ages. The average K isotopic composition of 8 enstatite chondrites is −0.47 ±0.57‰ (2SD, n=8), which is our best estimate for the bulk K isotopic composition of the enstatite chondrites. Compared to other extraterrestrial samples (*e.g.*, carbonaceous and ordinary chondrites, lunar, martian and eucrite meteorites), enstatite meteorites are the only group that have a K isotopic composition identical to that of the Earth (−0.48 ±0.03‰). Potassium stable isotopes are now added to the growing list of isotope systems which show indistinguishable compositions between enstatite meteorites and the Earth. Potassium isotopes of enstatite meteorites in this study further corroborate the isotopic similarity between Earth's building blocks and enstatite meteorite precursors.

We also observed the lightest K isotopic composition ($\delta^{41}K = -2.34 \pm 0.04‰$) in one aubrite fragment (MIL 13004) among all meteorites that have been analyzed to date. We attribute this very light K isotopic composition in this aubrite to the presence of larger amounts of djerfisherite in our sample because *ab initio* calculations predicted enrichments in lighter K isotopes. This conclusion needs to be test by further investigations.

## ACKNOWLEDGMENTS

We dedicate this manuscript to the memory of Dr. Christine Floss, who has studied enstatite meteorites during her PhD years at Washington University in St. Louis. We thank the Associate Editor Dr. Larry Nittler for prompt handling and thorough editing of this manuscript. We also thank reviewers Dr. Yogita Kadlag and Dr. Myriam Telus as well as one anonymous reviewer for their constructive comments. This work was supported by the McDonnell Center for the Space Sciences. Work by KL is supported in part by NSF AST 1517541. Work by MKP is supported by grant MSCA IPUSS-753276. We thank ANSMET, Dr. Kevin Righter and Meteorite Working Group at NASA's Johnson Space Center for collecting and curating Antarctic meteorites. US Antarctic meteorite samples are recovered by the Antarctic Search for Meteorites (ANSMET) program, which has been funded by NSF and NASA, and characterized and curated by the Department of Mineral Sciences of the Smithsonian Institution and Astromaterials Acquisition and Curation Office at NASA Johnson Space Center.

**Reference:**

Arevalo, R., McDonough, W.F. and Luong, M. (2009) The K/U ratio of the silicate Earth: Insights into mantle composition, structure and thermal evolution. *Earth and Planetary Science Letters* **278**, 361-369.

Baedecker, P.A. and Wasson, J.T. (1975) Elemental Fractionations among Enstatite Chondrites. *Geochimica et Cosmochimica Acta* **39**, 735-765.

Bland, P.A., Zolensky, M.E., Benedix, G.K. and Sephton, M.A. (2006) Weathering of Chondritic Meteorites. In: Lauretta, D.S., Jr., H.Y.M. (Eds.), Meteorites and the Early Solar System II, vol.943, pp.853-867, Tucson: University of Arizona Press

Bloom, H., Chen, H., Jr., B.F., Lodders, K. and Wang, K. (2018) Potassium Isotope Compositions of Carbonaceous and Ordinary Chondrites: Implications on the Origin of Volatile Depletion in the Early Solar System, *Lunar and Planetary Science Conference*. **49**, 1193

Boyet, M., Bouvier, A., Frossard, P., Hammouda, T., Garçon, M. and Gannoun, A. (2018) Enstatite chondrites EL3 as building blocks for the Earth: The debate over the 146Sm–142Nd systematics. *Earth and Planetary Science Letters* **488**, 68-78.

Brearley, A.J. and Jones, R.H. (1998) Chondritic meteorites. In: Papike, J.J. (Ed.), Planetary Materials, Reviews in Mineralogy, vol.36, pp.3-1–3-398, Washington, DC: Mineralogical Society of America

Burkhardt, C., Kleine, T., Oberli, F., Pack, A., Bourdon, B. and Wieler, R. (2011) Molybdenum isotope anomalies in meteorites: Constraints on solar nebula evolution and origin of the Earth. *Earth and Planetary Science Letters* **312**, 390-400.

Burnett, D.S., Lippolt, H.J. and Wasserburg, G.J. (1966) The relative isotopic abundance of K40 in terrestrial and meteoritic samples. *Journal of Geophysical Research (1896-1977)* **71**, 1249-1269.

Casanova, I., Keil, K. and Newsom, H.E. (1993) Composition of metal in aubrites: Constraints on core formation. *Geochimica et Cosmochimica Acta* **57**, 675-682.

Chen, H., Tian, Z., Tuller-Ross, B., Korotev, R.L. and Wang, K. (2019) High-precision potassium isotopic analysis by MC-ICP-MS: an inter-laboratory comparison and refined K atomic weight. *Journal of Analytical Atomic Spectrometry* **34**, 160-171.

Clayton, R.N., Mayeda, T.K. and Rubin, A.E. (1984) Oxygen isotopic compositions of enstatite chondrites and aubrites. *Journal of Geophysical Research* **89**, C245.

Crozaz, G. and Wadhwa, M. (2001) The terrestrial alteration of Saharan Shergottites Dar al Gani 476 and 489: A case study of weathering in a hot desert environment. *Geochimica et Cosmochimica Acta* **65**, 971-978.

Dauphas, N. (2017) The isotopic nature of the Earth's accreting material through time. *Nature* **541**, 521-524.

Dauphas, N., Chen, J.H., Zhang, J., Papanastassiou, D.A., Davis, A.M. and Travaglio, C. (2014) Calcium-48 isotopic anomalies in bulk chondrites and achondrites: Evidence for a uniform isotopic reservoir in the inner protoplanetary disk. *Earth and Planetary Science Letters* **407**, 96-108.

Dauphas, N., Marty, B. and Reisberg, L. (2002) Inference on terrestrial genesis from molybdenum isotope systematics. *Geophys Res Lett* **29**, 8-1-8-3.

Dauphas, N., Poitrasson, F., Burkhardt, C., Kobayashi, H. and Kurosawa, K. (2015) Planetary and meteoritic Mg/Si and $\delta^{30}$ Si variations inherited from solar nebula chemistry. *Earth and Planetary Science Letters* **427**, 236-248.

Defouilloy, C., Cartigny, P., Assayag, N., Moynier, F. and Barrat, J.A. (2016) High-precision sulfur isotope composition of enstatite meteorites and implications of the formation and evolution of their parent bodies. *Geochimica et Cosmochimica Acta* **172**, 393-409.

Ebel, D.S. and Sack, R.O. (2013) Djerfisherite: nebular source of refractory potassium. *Contributions to Mineralogy and Petrology* **166**, 923-934.

Ebel D. S., and Alexander C. M. O. (2011) Equilibrium condensation from chondritic porous IDP enriched vapor: Implications for Mercury and enstatite chondrite origins. *Planetary and Space Science* **59**. 1888–1894.

Fischer-Gödde, M., Burkhardt, C., Kruijer, T.S. and Kleine, T. (2015) Ru isotope heterogeneity in the solar protoplanetary disk. *Geochimica et Cosmochimica Acta* **168**, 151-171.

Fitoussi, C. and Bourdon, B. (2012) Silicon isotope evidence against an enstatite chondrite Earth. *Science* **335**, 1477-1480.

Fitoussi, C., Bourdon, B., Kleine, T., Oberli, F. and Reynolds, B.C. (2009) Si isotope systematics of meteorites and terrestrial peridotites: implications for Mg/Si fractionation in the solar nebula and for Si in the Earth's core. *Earth and Planetary Science Letters* **287**, 77-85.

Floss, C., Strait, M.M. and Crozaz, G. (1990) Rare earth elements and the petrogenesis of aubrites. *Geochimica et Cosmochimica Acta* **54**, 3553-3558.

Fogel, R.A., Hess, P.C. and Rutherford, M.J. (1989) Intensive parameters of enstatite chondrite metamorphism. *Geochimica et Cosmochimica Acta* **53**, 2735-2746.

Georg, R.B., Halliday, A.N., Schauble, E.A. and Reynolds, B.C. (2007) Silicon in the Earth's core. *Nature* **447**, 1102-1106.

Herzog, G., Albrecht, A., Ma, P., Fink, D., Klein, J., Middleton, R., Bogard, D.D., Nyquist, L., SHIH, C.Y. and Garrison, D. (2011) Cosmic - ray exposure history of the Norton County enstatite achondrite. *Meteoritics & Planetary Science* **46**, 284-310.

Herzog, G. and Caffee, M. (2014) Cosmic-ray exposure ages of meteorites. *Meteorites and cosmochemical processes*, 419-454.

Hidaka, H., Ebihara, M. and Yoneda, S. (1999) High fluences of neutrons determined from Sm and Gd isotopic compositions in aubrites. *Earth and Planetary Science Letters* **173**, 41-51.

Hu, Y., Chen, X.-Y., Xu, Y.-K. and Teng, F.-Z. (2018) High-precision analysis of potassium isotopes by HR-MC-ICPMS. *Chemical Geology* **493**, 100-108.

Huang, S. and Jacobsen, S.B. (2017) Calcium isotopic compositions of chondrites. *Geochimica et Cosmochimica Acta* **201**, 364-376.

Humayun, M. and Clayton, R.N. (1995) Potassium isotope cosmochemistry: Genetic implications of volatile element depletion. *Geochimica et Cosmochimica Acta* **59**, 2131-2148.

Javoy, M. (1995) The integral enstatite chondrite model of the Earth. *Geophys Res Lett* **22**, 2219-2222.

Javoy, M., Kaminski, E., Guyot, F., Andrault, D., Sanloup, C., Moreira, M., Labrosse, S., Jambon, A., Agrinier, P., Davaille, A. and Jaupart, C. (2010) The chemical composition of the Earth: Enstatite chondrite models. *Earth and Planetary Science Letters* **293**, 259-268.

Jull, A.J.T., Cloudt, S. and Cielaszyk, E. (1998) $^{14}C$ terrestrial ages of meteorites from Victoria Land, Antarctica, and the infall rates of meteorites. *Geological Society, London, Special Publications* **140**, 75.

Keil, K. (1989) Enstatite meteorites and their parent bodies*. *Meteoritics* **24**, 195-208.

Keil, K. (2010) Enstatite achondrite meteorites (aubrites) and the histories of their asteroidal parent bodies. *Chemie der Erde - Geochemistry* **70**, 295-317.

Keil, K., Mccoy, T.J., Wilson, L., Barrat, J.-A., Rumble, D., Meier, M.M.M., Wieler, R. and Huss, G.R. (2011) A composite Fe,Ni-FeS and enstatite-forsterite-diopside-glass vitrophyre clast in the Larkman Nunatak 04316 aubrite: Origin by pyroclastic volcanism. *Meteoritics & Planetary Science* **46**, 1719-1741.

Kempe, W. and Zähringer, J. (1966) K-Ar-Altersbestimmungen an Eisenmeteoriten—I Die Isotopenzusammensetzung des primären Kaliums. *Geochimica et Cosmochimica Acta* **30**, 1049-1057.

Kitts, K. and Lodders, K. (1998) Survey and evaluation of eucrite bulk compositions. *Meteoritics & Planetary Science* **33**, A197-A213.

Kong, P., Mori, T. and Ebihara, M. (1997) Compositional continuity of enstatite chondrites and implications for heterogeneous accretion of the enstatite chondrite parent body. *Geochimica et Cosmochimica Acta* **61**, 4895-4914.

Krot, A.N., Keil, K., Scott, E.R.D., Goodrich, C.A. and Weisberg, M.K. (2014) 1.1 - Classification of Meteorites and Their Genetic Relationships. In: Holland, H.D., Turekian, K.K. (Eds.), Treatise on Geochemistry (Second Edition), pp.1-63, Oxford: Elsevier

Ku, Y. and Jacobsen, S.B. (2019) Potassium isotope variation in chondrites, *Lunar and Planetary Science Conference*. **50**, 1675

Larimer, J.W. and Anders, E. (1970) Chemical fractionations in meteorites: III. Major element fractionations in chondrites. *Geochimica et Cosmochlmica Acta* **34**, 367-387.

Lee, J.Y., Marti, K. and Wacker, J.F. (2009) Xe isotopic abundances in enstatite meteorites and relations to other planetary reservoirs. *Journal of Geophysical Research* **114**.

Li, W., Beard, B.L. and Li, S. (2016) Precise measurement of stable potassium isotope ratios using a single focusing collision cell multi-collector ICP-MS. *Journal of Analytical Atomic Spectrometry* **31**, 1023-1029.

Li, Y., Wang, W. and Zhongqing, W. (in review) First-principles investigation of equilibrium K isotope fractionation among K-bearing minerals. *submited to GCA*.

Lipschutz, M.E., Verkouteren, R.M., Sears, D.W.G., Hasan, F.A., Prinz, M., Weisberg, M.K., Nehru, C.E., Delaney, J.S., Grossman, L. and Boily, M. (1988) Cumberland

Falls chondritic inclusions: III. Consortium study of relationship to inclusions in Allan Hills 78113 aubrite. *Geochimica et Cosmochimica Acta* **52**, 1835-1848.

Lodders, K. (2003) Solar system abundances and condensation temperatures of the elements. *The Astrophysical Journal* **591**, 1220-1247.

Lodders, K. and Fegley, B. (1998) The planetary scientist's companion. Oxford University Press on Demand.

Lodders, K., Palme, H. and Wlotzka, F. (1993) Trace elements in mineral separates ofthe Peiia Blanca Spring aubrite: Implications for the evolution of the aubrite parent body. *Meteoritics* **28**, 538-551.

Lorenzetti, S., Eugster, O., Busemann, H., Marti, K., Burbine, T.H. and McCoy, T. (2003) History and origin of aubrites. *Geochimica et Cosmochimica Acta* **67**, 557-571.

Meija, J., Coplen Tyler, B., Berglund, M., Brand Willi, A., De Bièvre, P., Gröning, M., Holden Norman, E., Irrgeher, J., Loss Robert, D., Walczyk, T. and Prohaska, T. (2016) Atomic weights of the elements 2013 (IUPAC Technical Report), *Pure and Applied Chemistry*, p. 265.

Morgan, L.E., Santiago Ramos, D.P., Davidheiser-Kroll, B., Faithfull, J., Lloyd, N.S., Ellam, R.M. and Higgins, J.A. (2018) High-precision 41K/39K measurements by MC-ICP-MS indicate terrestrial variability of δ41K. *Journal of Analytical Atomic Spectrometry* **33**, 175-186.

Mougel, B., Moynier, F. and Göpel, C. (2018) Chromium isotopic homogeneity between the Moon, the Earth, and enstatite chondrites. *Earth and Planetary Science Letters* **481**, 1-8.

Moynier, F., Paniello, R.C., Gounelle, M., Albarède, F., Beck, P., Podosek, F. and Zanda, B. (2011) Nature of volatile depletion and genetic relationships in enstatite chondrites and aubrites inferred from Zn isotopes. *Geochimica et Cosmochimica Acta* **75**, 297-307.

Nishiizumi, K., Elmore, D. and Kubik, P.W. (1989) Update on Terrestrial Ages of Antarctic Meteorites. *Earth and Planetary Science Letters* **93**, 299-313.

Ntaflos, T. and Koeberl, C. (1992) Petrological studies and bulk chemical analyses of eight antarctic aubrites, *Lunar and Planetary Science Conference*. **23**, 1003

Parendo, C.A., Jacobsen, S.B. and Wang, K. (2017) K isotopes as a tracer of seafloor hydrothermal alteration. *Proc Natl Acad Sci U S A* **114**, 1827-1831.

Patzer, A. and Schultz, L. (2001) Noble gases in enstatite chondrites I: Exposure ages, pairing, and weathering effects. *Meteoritics & Planetary Science* **36**, 947-961.

Petaev, M.I. and Khodakovsky, I.L. (1986) Thermodynamic Properties and Conditions of Formation of Minerals in Enstatite Meteorite, *Chemistry and Physics of Terrestrial Planets*. 106-135

Pringle, E.A. and Moynier, F. (2017) Rubidium isotopic composition of the Earth, meteorites, and the Moon: Evidence for the origin of volatile loss during planetary accretion. *Earth and Planetary Science Letters* **473**, 62-70.

Qin, L., Alexander, C.M.O.D., Carlson, R.W., Horan, M.F. and Yokoyama, T. (2010) Contributors to chromium isotope variation of meteorites. *Geochimica et Cosmochimica Acta* **74**, 1122-1145.

Rubin, A.E. (2010) Impact melting in the Cumberland Falls and Mayo Belwa aubrites. *Meteoritics & Planetary Science* **45**, 265-275.

Rubin, A.E. (2015) Impact features of enstatite-rich meteorites. *Chemie der Erde - Geochemistry* **75**, 1-28.

Rubin, A.E., Huber, H. and Wasson, J.T. (2009) Possible impact-induced refractory-lithophile fractionations in EL chondrites. *Geochimica et Cosmochimica Acta* **73**, 1523-1537.

Rubin, A.E., Scott, E.R.D. and Keil, K. (1997) Shock metamorphism of enstatite chondrites. *Geochimica et Cosmochimica Acta* **61**, 847-858.

Rubin, A.E. and Wasson, J.T. (2011) Shock effects in “EH6” enstatite chondrites and implications for collisional heating of the EH and EL parent asteroids. *Geochimica et Cosmochimica Acta* **75**, 3757-3780.

Santiago Ramos, D.P., Morgan, L.E., Lloyd, N.S. and Higgins, J.A. (2018) Reverse weathering in marine sediments and the geochemical cycle of potassium in seawater: Insights from the K isotopic composition ( $^{41}K/^{39}K$) of deep-sea pore-fluids. *Geochimica et Cosmochimica Acta* **236**, 99-120.

Savage, P.S. and Moynier, F. (2013) Silicon isotopic variation in enstatite meteorites: Clues to their origin and Earth-forming material. *Earth and Planetary Science Letters* **361**, 487-496.

Schauble, E.A. (2004) Applying Stable Isotope Fractionation Theory to New Systems. *Reviews in Mineralogy & Geochemistry* **55**, 65-111.

Steele, R.C.J., Coath, C.D., Regelous, M., Russell, S. and Elliott, T. (2012) Neutron-poor Nickel Isotope Anomalies in Meteorites. *The Astrophysical Journal* **758**, 59.

Tang, H. and Dauphas, N. (2012) Abundance, distribution, and origin of 60Fe in the solar protoplanetary disk. *Earth and Planetary Science Letters* **359-360**, 248-263.

Teng, F.-Z., Li, W.-Y., Ke, S., Marty, B., Dauphas, N., Huang, S., Wu, F.-Y. and Pourmand, A. (2010) Magnesium isotopic composition of the Earth and chondrites. *Geochimica et Cosmochimica Acta* **74**, 4150-4166.

Tian, Z., Chen, H., B. Fegley, J., Lodders, K., Barrat, J.-A. and Wang, K. (2018) Potassium Isotope Differences among Chondrites, Earth, Moon, Mars, and 4-Vesta - Implication on the Planet Accretion Mechanisms, *Lunar and Planetary Science Conference*. **49**, 1276

Trinquier, A., Birck, J. and Allegre, C.J. (2007) Widespread 54Cr Heterogeneity in the Inner Solar System. *The Astrophysical Journal* **655**, 1179.

Trinquier, A., Elliott, T., Ulfbeck, D., Coath, C., Krot, A.N. and Bizzarro, M. (2009) Origin of nucleosynthetic isotope heterogeneity in the solar protoplanetary disk. *Science* **324**, 374-376.

Udry, A., Wilbur, Z.E., Rahib, R.R., McCubbin, F.M., Vander Kaaden, K.E., McCoy, T.J., Ziegler, K., Gross, J., DeFelice, C., Combs, L. and Turrin, B.D. (2019) Reclassification of four aubrites as enstatite chondrite impact melts: Potential geochemical analogs for Mercury. *Meteoritics & Planetary Science* **54**, 785-810.

Velbel, M.A., Long, D.T. and Gooding, J.L. (1991) Terrestrial weathering of Antarctic stone meteorites: Formation of Mg-carbonates on ordinary chondrites. *Geochimica et Cosmochimica Acta* **55**, 67-76.

Voshage, H., Feldmann, H. and Braun, O. (1983) Investigations of cosmic-ray-produced nuclides in iron meteorites: 5. More data on the nuclides of potassium and noble gases, on exposure ages and meteoroid sizes. *Zeitschrift für Naturforschung A* **38**, 273-280.

Wang, K. and Jacobsen, S.B. (2016a) An estimate of the Bulk Silicate Earth potassium isotopic composition based on MC-ICPMS measurements of basalts. *Geochimica et Cosmochimica Acta* **178**, 223-232.

Wang, K. and Jacobsen, S.B. (2016b) Potassium isotopic evidence for a high-energy giant impact origin of the Moon. *Nature* **538**, 487-490.

Wang, K., Savage, P.S. and Moynier, F. (2014) The iron isotope composition of enstatite meteorites: Implications for their origin and the metal/sulfide Fe isotopic fractionation factor. *Geochimica et Cosmochimica Acta* **142**, 149-165.

Watters, T.R. and Prinz, M. (1979) Aubrites: their origin and relationship to enstatite chondrites, *Lunar and Planetary Science Conference*. **10**, 1073-1093

Welten, K., Nishiizumi, K., Hillegonds, D., Caffee, M. and Masarik, J. (2004) Unraveling the exposure histories of aubrites. *Meteoritics & Planetary Science* **39**, #5220.

Wolf, R., Ebihara, M. and R.Richter, G. (1983) Aubrites and diogenites: Trace element clues to their origin. *Geochimica et Cosmochimica Acta* **47**, 14.

Wombacher, F., Rehkämper, M., Mezger, K., Bischoff, A. and Münker, C. (2008) Cadmium stable isotope cosmochemistry. *Geochimica et Cosmochimica Acta* **72**, 646-667.

Wombacher, F., Rehkämper, M., Mezger, K. and Münker, C. (2003) Stable isotope compositions of cadmium in geological materials and meteorites determined by multiple-collector ICPMS. *Geochimica et Cosmochimica Acta* **67**, 4639-4654.

Yanai, K., Kojima, H. and H, H. (1995) Catalog of the Antartic meteorites collected from December 1969 to December 1994, with special reference to those represented in the collections of the National Institute of Polar Research. pp.44-76, 212-213, Tokyo: National Institut of Polar Research

Zhang, J., Dauphas, N., Davis, A.M., Leya, I. and Fedkin, A. (2012) The proto-Earth as a significant source of lunar material. *Nature Geoscience* **5**, 251-255.

Zhang, Y., Benoit, P.H. and Sears, D.W.G. (1995) The classification and complex thermal history of the enstatite chondrites. *Journal of Geophysical Research: Planets* **100**, 9417-9438.

**Table 1. The chemical classification, petrological types, K concentration and isotopic compositions of enstatite chondrites in this study.**

| Name | Type | Weathering[a] | Shock[a] | Mass (mg) | K (ppm)[e] | $\delta^{41}K$ | ±2σ | n |
|---|---|---|---|---|---|---|---|---|
| **MAC 88136** | **EL3** | **A** | **S3** | 82.2 | 565 | **−0.740** | **0.045** | **12** |
| **MAC 02747 Ave.** [c] | **EL4** | **B/C** | **S4** | | | **−0.756** | **0.020** | **23** |
| *MAC 02747 #1*[b] | | | | *118.7* | *616* | *−0.771* | *0.031* | *13* |
| *MAC 02747 #2*[b] | | | | *54.8* | *256* | *−0.744* | *0.027* | *10* |
| **ALHA81021** | **EL6** | **A** | **S2** | 188.3 | 425 | **−0.918** | **0.040** | **13** |
| **LON 94100 Ave.** [c] | **EL6** | **A/Be** | **S2** | | | **−0.180** | **0.028** | **25** |
| *LON 94100 #1*[b] | | | | *125.3* | *518* | *−0.194* | *0.036* | *13* |
| *LON 94100 #2*[b] | | | | *33.0* | *471* | *−0.157* | *0.046* | *12* |
| **Average EL** | | | | | | **−0.625±0.645**[d] | | |
| **MIL 07028 Ave.** [c] | **EH3** | **B** | | | | **−0.349** | **0.020** | **23** |
| *MIL 07028 #1*[b] | | | | *73.6* | *604* | *−0.418* | *0.038* | *13* |
| *MIL 07028 #2*[b] | | | | *36.9* | *668* | *−0.322* | *0.024* | *10* |
| **LAR 12252 Ave.** [c] | **EH3** | **A** | | | | **−0.180** | **0.029** | **22** |
| *LAR 12252 #1*[b] | | | | *112.9* | *220* | *−0.196* | *0.054* | *10* |
| *LAR 12252 #2*[b] | | | | *37.4* | *215* | *−0.173* | *0.034* | *12* |
| **LAR 06252 Ave.** [c] | **EH3** | **Be** | **S4** | | | **−0.610** | **0.021** | **21** |
| *LAR 06252 #1*[b] | | | | *124.4* | *498* | *−0.626* | *0.027* | *11* |
| *LAR 06252 #2*[b] | | | | *67.4* | *588* | *−0.582* | *0.035* | *10* |
| **EET 96135 Ave.**[c] | **EH4/5** | **B** | **S3** | | | **−0.325** | **0.020** | **35** |
| *EET 96135 #1*[b] | | | | *99.1* | *575* | *−0.278* | *0.033* | *11* |
| *EET 96135 #2*[b] | | | | *54.6* | *748* | *−0.407* | *0.042* | *13* |
| *EET 96135 #3*[b] | | | | | | *−0.323* | *0.031* | *11* |
| **Average EH** | | | | | | **−0.386±0.358**[d] | | |
| **Average All** | | | | | | **−0.473±0.569**[d] | | |
| BHVO-2 | Geostandard | | | | | −0.463 | 0.031 | 91 |

[a] The weathering and shock conditions are from MetBase and references therein (Zhang et al., 1995; Kong et al., 1997; Rubin et al., 2009; Wang et al., 2014; Rubin, 2015). Weathering index A, B or C represents "minor", "moderate" or "severe" rustiness, respectively. Letter "e" represents evaporite minerals visible.

[b] The "#" represents repeated sample from the same powder of the same sample.

[c] The average value of a sample was reported with error-weighted average of all measurements and the uncertainty is given with ±2σ.

[d] The average value of the group was reported with error-weighted average of all meteorites in the group and the uncertainty was given with 2SD.

[e] The uncertainty of the K concentration is 10% RSD based on the repeated measurements using iCapQ quadrupole ICP-MS.

**Table 2. Petrological type, K concentration and isotopic compositions of aubrites in this study.**

| Name | Type | Weathering[a] | Shock[a] | Mass (mg) | K (ppm)[d] | $\delta^{41}K$ | ±2σ | n |
|---|---|---|---|---|---|---|---|---|
| **ALHA78113 Ave.[c]** | **AUB** | **A/Be** | **S2-5** | | | **−0.632** | **0.035** | **20** |
| *ALHA78113 fragment 1[b]* | | | | *177.30* | *62* | *−0.629* | *0.051* | *10* |
| *ALHA78113 fragment 2[b]* | | | | *145.8* | *49* | *−0.634* | *0.048* | *10* |
| **ALH 84009** | **AUB** | **A** | | 283.90 | 90 | **−0.752** | **0.042** | **13** |
| **LAR 04316 Ave.[c]** | **AUB** | **A** | **S4** | | | **−0.374** | **0.032** | **21** |
| *LAR 04316 fragment 1[b]* | | | | *216.10* | *240* | *−0.338* | *0.035* | *11* |
| *LAR 04316 fragment 2[b]* | | | | *122.5* | *217* | *−0.542* | *0.075* | *10* |
| **MIL 13004 Ave.[c]** | **AUB** | **A/B** | | | | **−2.341** | **0.041** | **22** |
| *MIL 13004 fragment 1[b]* | | | | *104.50* | *313* | *−2.339* | *0.043* | *11* |
| *MIL 13004 fragment 2[b]* | | | | *115.5* | *340* | *−2.358* | *0.124* | *11* |

[a] Weathering and shock degrees are from MetBase and references therein (Lipschutz et al., 1988; Ntaflos and Koeberl, 1992; Yanai et al., 1995; Keil et al., 2011; Wang et al., 2014; Rubin, 2015). Weathering index A, B or C represents "minor", "moderate" or "severe" rustiness, respectively. Letter "e" represents evaporite minerals visible.

[b] Different fragments of the same meteorite were analyzed.

[c] The average value of a sample was reported with error-weighted average of all measurements and the uncertainty is given with ±2σ.

[d] The uncertainty of the K concentration is 10% RSD based on the repeated measurements using iCapQ quadrupole ICP-MS.

## Figure Captions

**Fig. 1.** Potassium isotope ($\delta^{41}K$) variations of different subgroups of enstatite meteorites. Error bars shown are 2σ (95% confidence intervals) and are typically smaller than the symbol sizes. The dashed line stands for the average $\delta^{41}K$ value of enstatite chondrites ($\delta^{41}K = -0.47$). The shaded bar represents the Bulk Silicate Earth $\delta^{41}K = -0.48 \pm 0.03$ ‰ (Wang and Jacobsen, 2016a).

**Fig. 2**. Comparison of data obtained for aliquots of the samples repeatedly analyzed in this study.

**Fig. 3**. Range of $\delta^{41}K$ observed for undifferentiated (OC, CC and EC) and differentiated (Moon, Vesta, Mars, and aubrites) meteorites in this study and from the literature (Tian et al., 2018; Tian et al., in review; Bloom et al., 2018; Wang and Jacobson, 2016b; Ku and Jacobsen, 2019).

**Fig. 4**. Histograms of K isotope compositions ($\delta^{41}K$) for carbonaceous, enstatite, and ordinary chondrites as well as aubrites. The outlier, aubrite MIL 13004, is excluded from this figure. All data shown are a combination of this study and the literature (Bloom et al., 2018; Wang and Jacobson, 2016b; Ku and Jacobsen, 2019).

**Fig. 5**. The K isotopic compositions of enstatite meteorites as functions of (a) weathering degree (b) petrological type (c) shock degree.

**Fig. 6**. The K isotopic compositions of enstatite meteorites as functions of (a) Fe/K ratio (b) Ca/K ratio (c) cosmic-ray exposure ages. The data are from this study

and from the literature (Lipschutz et al., 1988; Ntaflos and Koeberl, 1992; Yanai et al., 1995; Zhang et al., 1995; Kong et al., 1997; Patzer and Schultz, 2001; Lorenzetti et al., 2003; Rubin et al., 2009; Keil et al., 2011; Wang et al., 2014).

**Fig. 7**. $\delta^{41}$K vs. the inverse potassium content [K] (ppm), for enstatite chondrites and aubrites analyzed in this study.

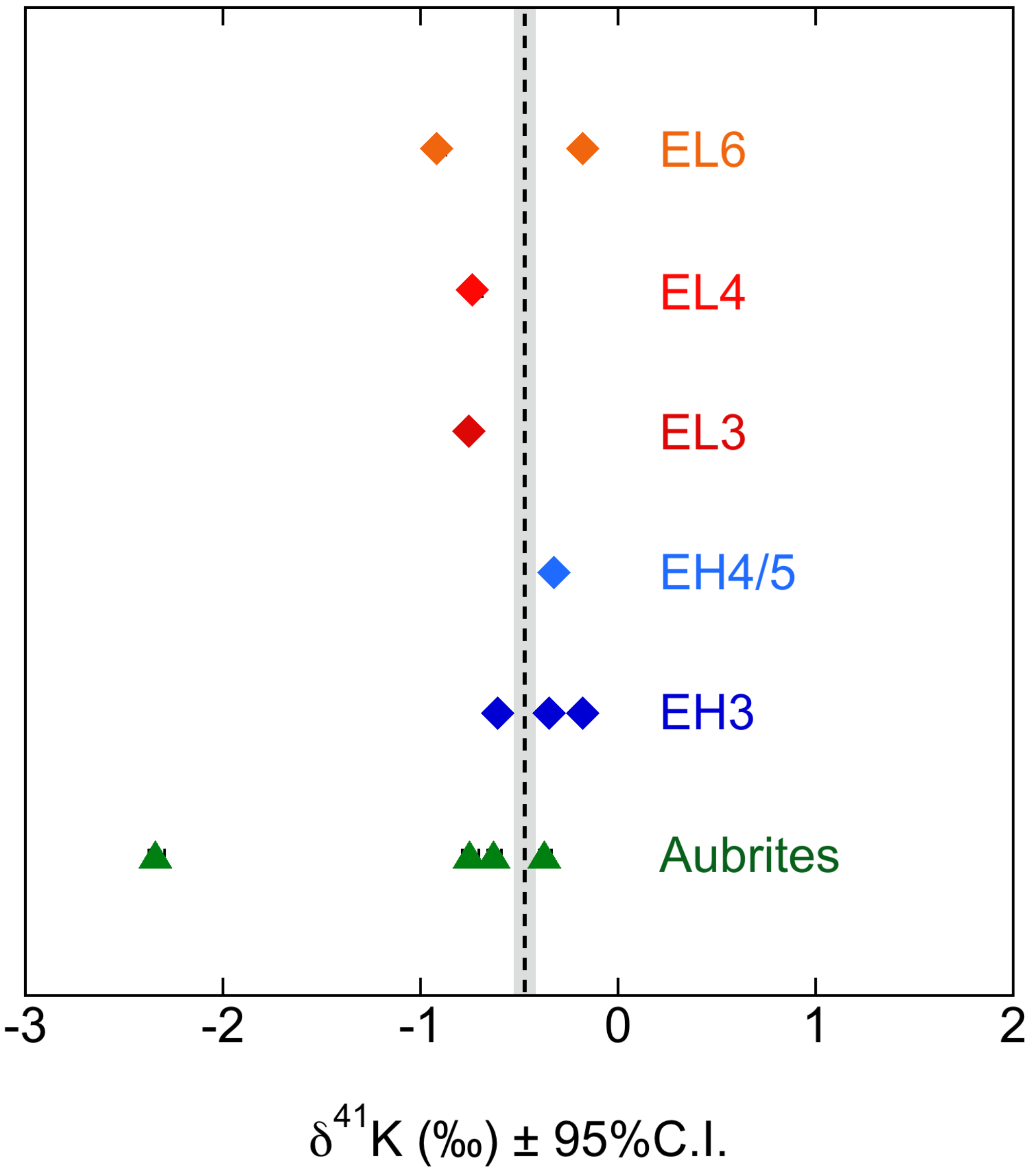

Fig.1

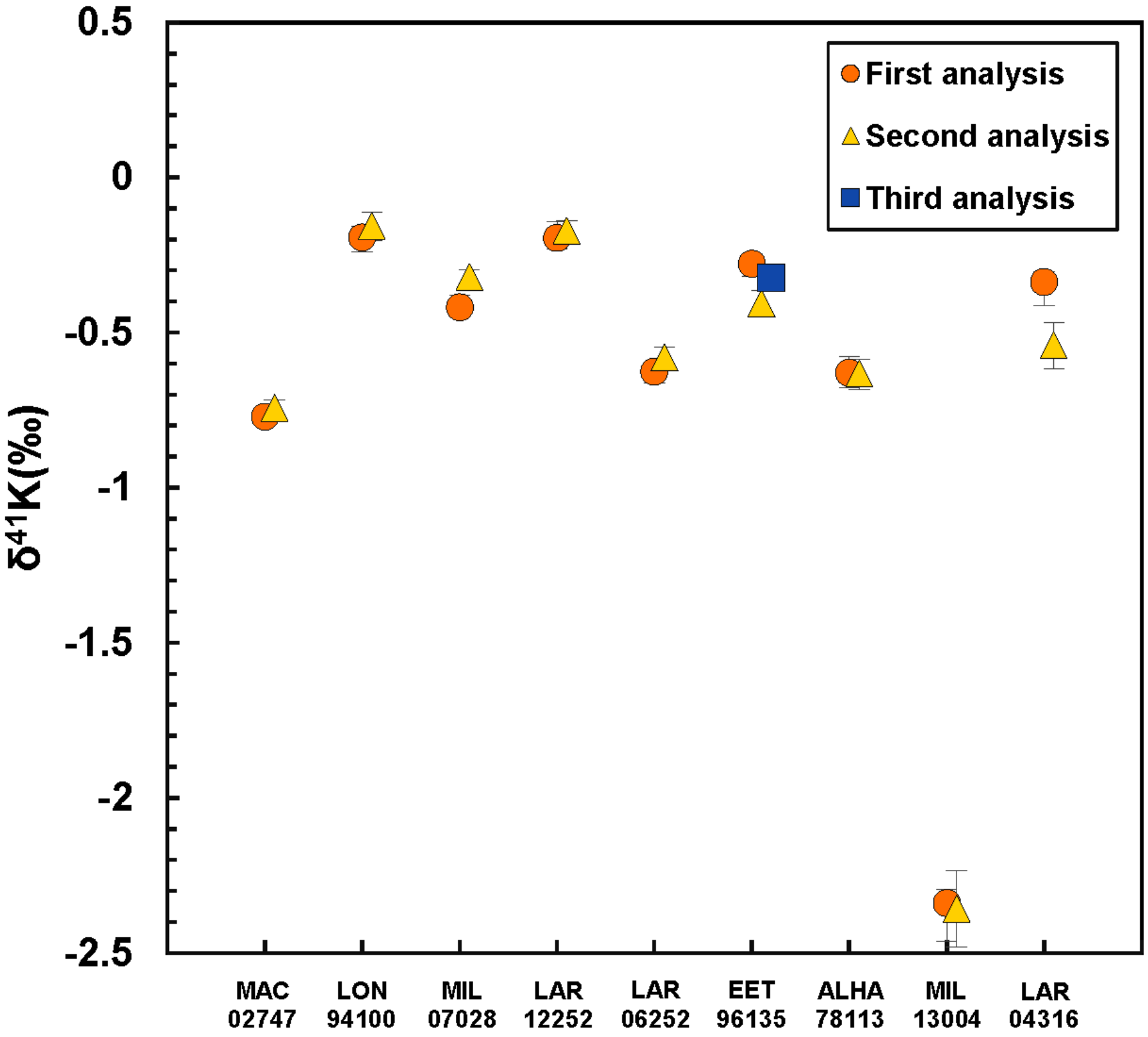

Fig.2

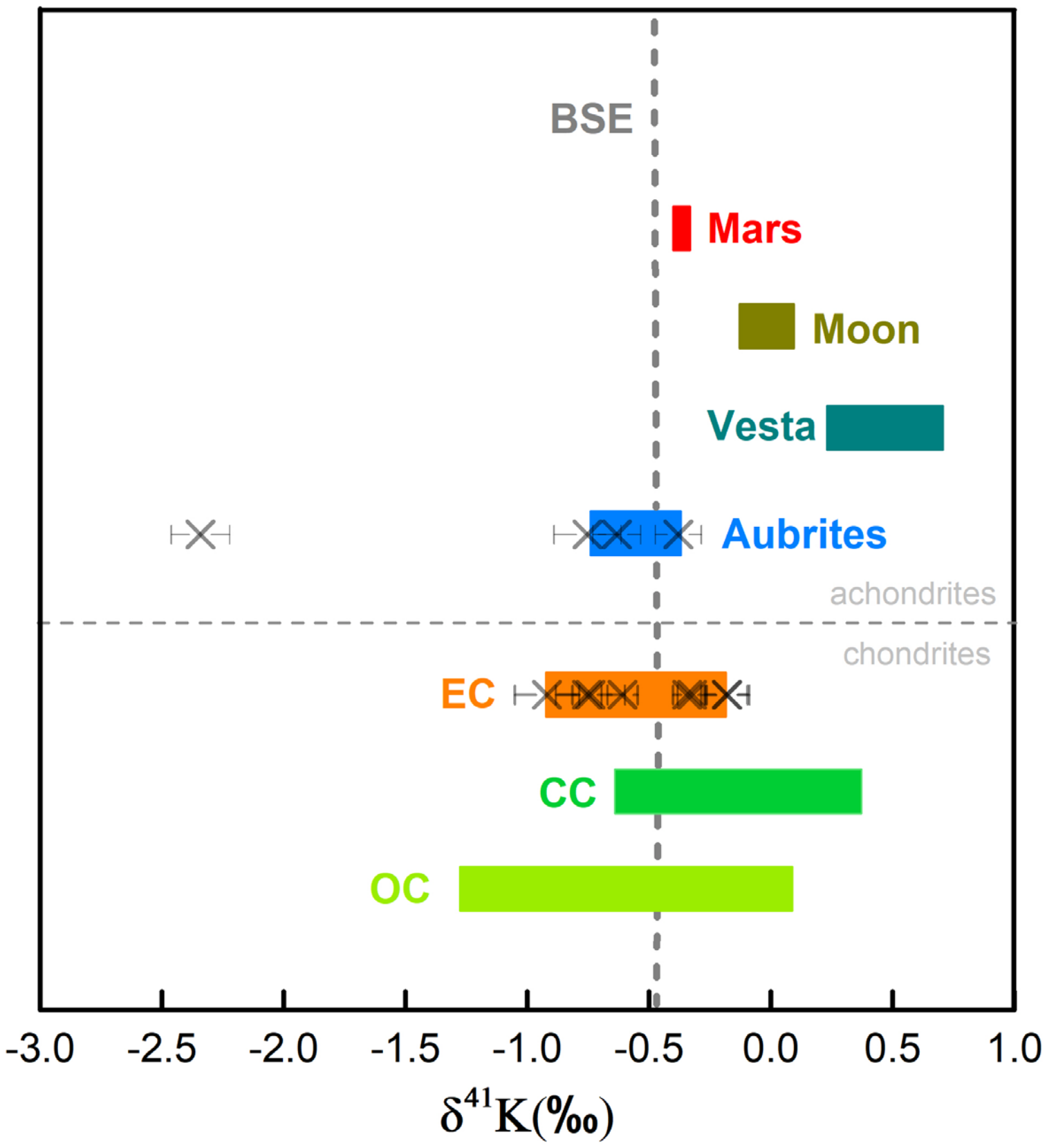

65 Fig.3

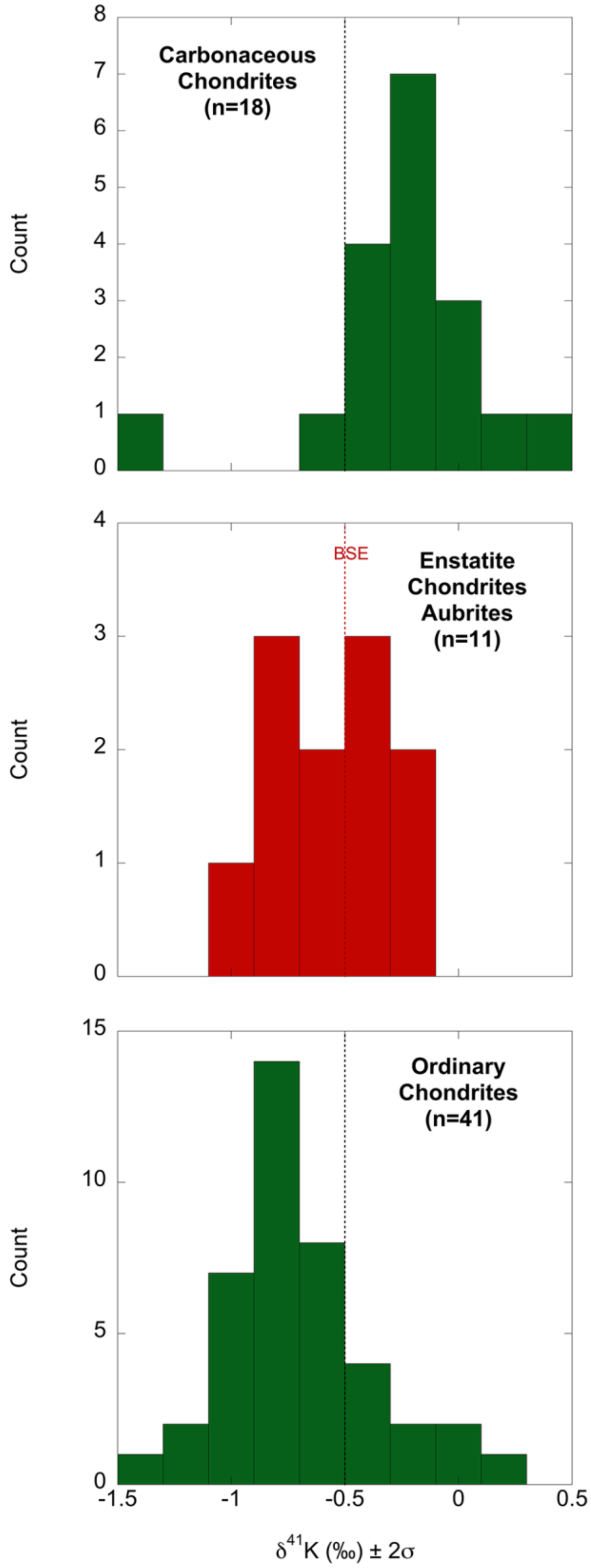

Fig. 4

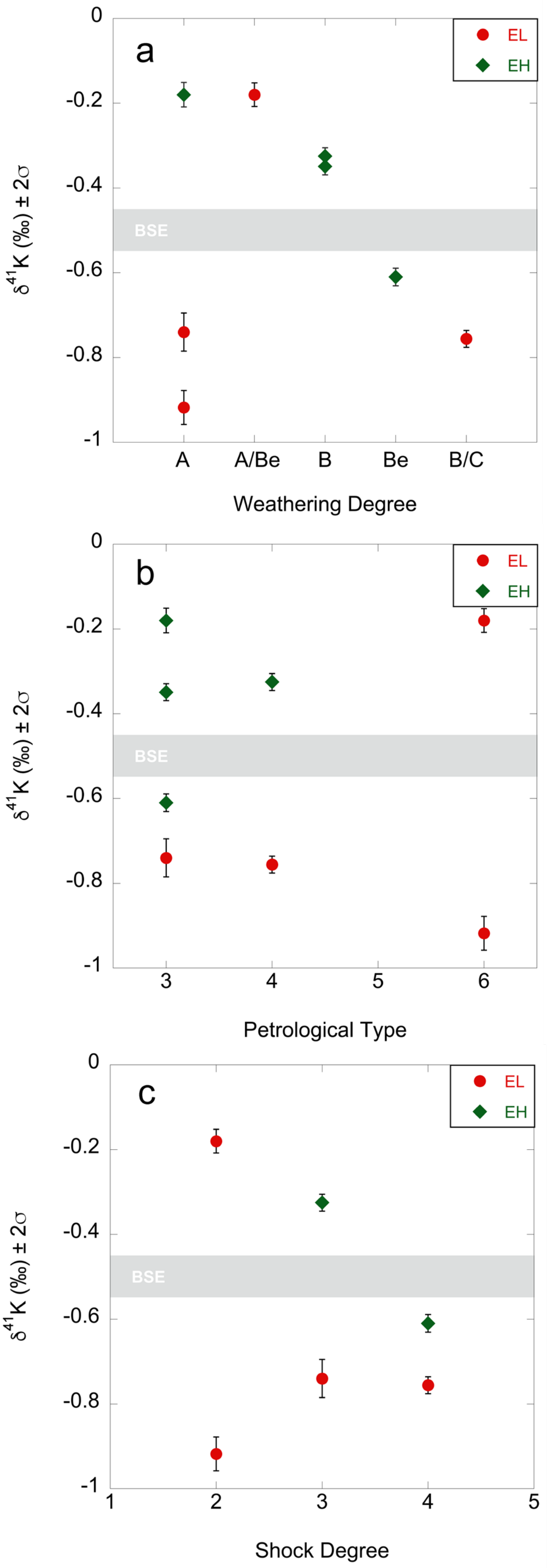

Fig. 5

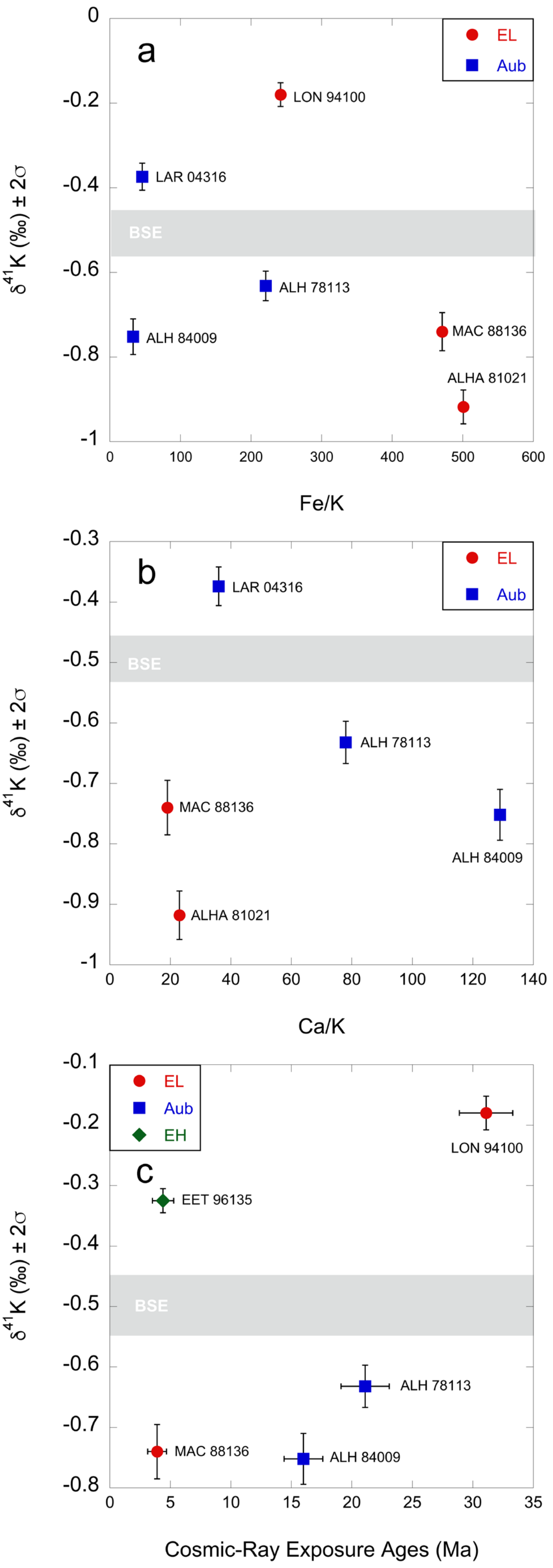

Fig. 6

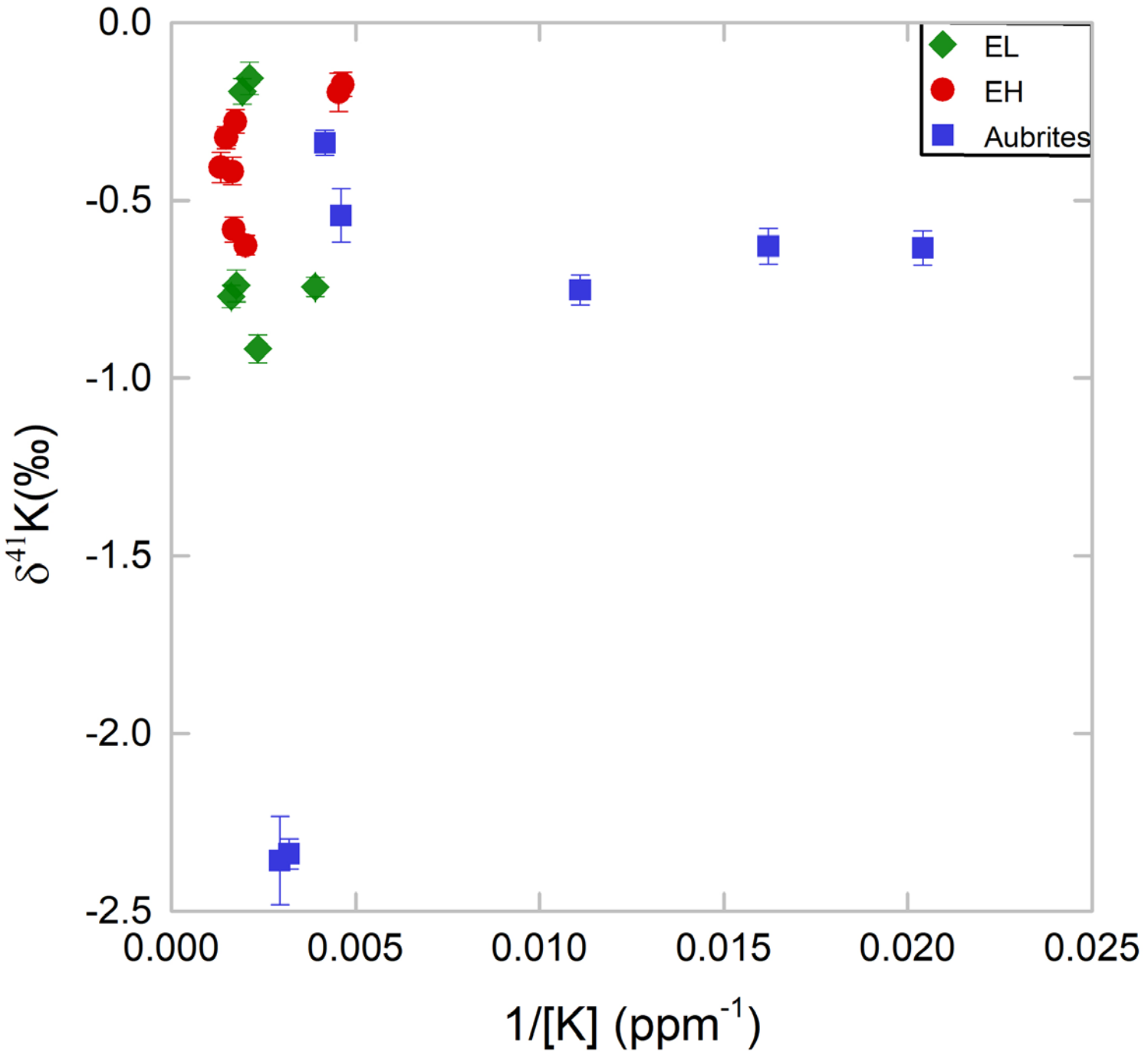

Fig. 7